\documentclass[sigconf]{acmart}

\usepackage[yyyymmdd]{datetime}

\newcommand{\timestr}{\today}

\setcopyright{none} 
\acmConference[Anonymous Submission to ACM CCS 2017]{ACM Conference on Computer and Communications Security}{Due 19 May 2017}{Dallas, Texas}
\acmYear{2017}

\usepackage{amssymb, amsmath, amsthm}
\usepackage{listings}
\usepackage{tabularx}
\usepackage{enumerate}
\usepackage{semantic}

\usepackage{lstlangcoq}

\newcommand{\langC}{\lstset{language=C,{morekeywords={bool}},basicstyle={\ttfamily},columns=fullflexible,literate=}}
\newcommand{\langCoq}{\lstset{language=Coq,basicstyle=\ttfamily,mathescape=true,columns=fullflexible}}

\usepackage{makecell}
\usepackage{array}

\langCoq

\usepackage{tikz}
\usetikzlibrary{positioning, calc}
\usetikzlibrary{shapes}
\usetikzlibrary{arrows.meta}
\usetikzlibrary{fit}

\newcommand{\ap}[2]{#1~#2}
\newcommand{\app}[3]{#1~#2~#3}
\newcommand{\appp}[4]{#1~#2~#3~#4}

\newcommand{\EK}[1]{\textsf{EK}_{\textsf{#1}}}
\newcommand{\EKnormal}{\EK{nrm}}
\newcommand{\EKbreak}{\EK{brk}}
\newcommand{\EKcontinue}{\EK{cont}}
\newcommand{\EKreturn}{\EK{ret}}

\newcommand{\excont}{\mbox{\lstinline|exit_cont|}}

\newcommand{\None}{\textsf{None}}

\newcommand{\selek}[5]{(#1,\, #2,\, #3,\, #4)_{#5}}
\newcommand{\selekv}[5]{\left(\begin{matrix}#1\\ #2\\ #3\\ #4\end{matrix}\right)_{#5}}

\newcommand{\fun}[2]{\lambda #1.\,#2}
\newcommand{\funtpd}[3]{\lambda #1:#2.\,#3}

\newcommand{\ite}[3]{\textbf{if}~#1~\textbf{then}~#2~\textbf{else}~#3}
\newcommand{\mathfalse}{\textsf{false}}
\newcommand{\mathtrue}{\textsf{true}}
\newcommand{\mathcoq}[1]{\textsf{#1}}

\newcommand{\nil}{\mathcoq{nil}}

\newcommand{\Lo}{\mathcoq{Lo}}
\newcommand{\Hi}{\mathcoq{Hi}}

\newcommand{\lub}[2]{#1 \sqcup #2}
\newcommand{\glb}[2]{#1 \sqcap #2}
\newcommand{\lle}[2]{#1 \sqsubseteq #2}

\newcommand{\triple}[3]{\{#1\}\,#2\,\{#3\}}
\newcommand{\vsttriple}[4]{#1 \vdash \triple{#2}{#3}{#4}}
\newcommand{\ifcdef}[4]{\vsttriple{#1}{#2}{#3}{#4}}

\newcommand{\trp}[1]{\mathcal{#1}}

\newcommand{\gfsvec}[1]{#1}

\newcommand{\fieldaddress}[2]{#1 \triangleright #2}
\newcommand{\fieldat}[2]{ \shortmid \hspace{-0.42ex} \xrightarrow[{#1}]{#2}}
\newcommand{\concatpath}[2]{#1 \mbox{\textsf{++}} #2}

\newcommand{\nret}[1]{\mathcoq{nret}~#1}

\newcommand{\cident}[1]{\texttt{#1}}
\newcommand{\ckeyword}[1]{\texttt{\textbf{#1}}}
\newcommand{\cloop}[2]{\ckeyword{loop}(#1)#2}
\newcommand{\cbreak}{\ckeyword{break}}
\newcommand{\ccontinue}{\ckeyword{continue}}
\newcommand{\creturnNonVoid}[1]{\ckeyword{return}~#1}
\newcommand{\cif}[3]{\ckeyword{if}~#1~\ckeyword{then}~#2~\ckeyword{else}~#3}

\newcommand{\denote}[1]{\llbracket #1 \rrbracket}

\newcommand{\PROP}{\textsc{prop}}
\newcommand{\LOCAL}{\textsc{local}}
\newcommand{\SEP}{\textsc{sep}}
\newcommand{\PLS}[3]{\PROP\,#1\,\LOCAL\,#2\,\SEP\,#3}
\newcommand{\ltemp}[2]{\mathsf{temp}~\mathsf{#1}~#2}

\newcommand{\clstep}[2]{#1 \rightarrow #2}
\newcommand{\clstar}[2]{#1 \rightarrow^\ast #2}
\newcommand{\clstarN}[3]{#1 \rightarrow_{#2} #3}
\newcommand{\clbigstep}[2]{#1 \Downarrow #2}

\newcommand{\clstate}[3]{\langle #1, #2, #3 \rangle}

\newcommand{\specialequiv}[1]{\equiv_{\textsf{#1}}}
\newcommand{\contequiv}[2]{#1 \specialequiv{cont} #2}
\newcommand{\cscontheadequiv}[2]{#1 \specialequiv{head} #2}
\newcommand{\sync}[2]{#1 \specialequiv{sync} #2}
\newcommand{\guardkw}[1]{}
\newcommand{\genguard}[3]{\guardkw{#1}\{#2\}~#3}
\newcommand{\guard}[2]{\genguard{guard}{#1}{#2}}
\newcommand{\rguard}[2]{\genguard{rguard}{#1}{#2}}
\newcommand{\iguard}[3]{\genguard{iguard}{#1}{#2}~#3}
\newcommand{\irguard}[3]{\genguard{irguard}{#1}{#2}~#3}

\begin{document}

\title{VST-Flow: Fine-grained low-level \\ reasoning about real-world C code (Draft)}
\author{Samuel Gruetter \and Toby Murray}

\begin{abstract}
We show how support for information-flow security proofs could be added on top of the Verified Software Toolchain (VST).
We discuss several attempts to define information flow security in a VST-compatible way, and present a statement of information flow security in ``continuation-passing'' style.

Moreover, we present Hoare rules augmented with information flow control assertions, and sketch how these rules could be proven sound with respect to the definition given before.

We also discuss how this can be implemented in the Coq proof assistant,\footnote{The Coq development is available at \url{https://github.com/samuelgruetter/vst-ifc}.} and how VST's proof automation framework (VST-Floyd) can be adapted to support convenient information flow security proofs.
\end{abstract}

%



\maketitle

\section{Introduction}



Software plays a crucial role in today's world, and software bugs can cause great damage.
To minimize the impact of bugs and malicious behavior, various kinds of ``supervisor'' and information dispatch software has been developed, such as operating systems, hypervisors, network switches, firewalls, etc, which manage and restrict the interactions between different software components to achieve certain security guarantees.

For performance reasons, the interaction managing software is often implemented in low-level languages such as C, and it is hard to ensure that these crucial pieces of software are correct.
The Verified Software Toolchain (VST) \cite{VST} offers a way to formally verify functional correctness of C programs, but it does not support reasoning about the information flow between different components.

Therefore, we present VST-Flow, a proposal for an extension to VST allowing to reason about information flow policies.
That is, we introduce a way to assign a high or low sensitivity level to data, and extend VST's Hoare rules to include these classifications, which results in a convenient technique to prove non-interference for C programs, i.e. to prove that the values of high data do not influence the values of low data.

Many such systems have been presented before \cite{zheng2007,costanzo2014,murray2016}, where the user can prove that a program respects a desired information flow policy by applying a given set of rules.
The judgment form of these rules might vary between different approaches, some being closer to typing rules, while others being closer to Hoare rules, but as far as we know, the soundness statement of all these systems is in ``direct-style''.
By ``direct-style'', we mean that the soundness statement roughly has the form ``\emph{if some condition $\phi_1$ holds in state $\sigma_1$, and we run the command $c$, then we end up in a state $\sigma_2$ where the condition $\phi_2$ holds}.''

Depending on the approach, $\phi_1$ and $\phi_2$ might be the same, and in the typical case, $\sigma_1$ and $\sigma_2$ are pairs of two states, and $\phi_1$ and $\phi_2$ state low-equivalence, i.e. that all low values have the same value in the two states of the pair.

However, as explained by Appel \cite{appel2007,appel2014}, to support premature exits such as \lstinline|break|, \lstinline|continue|, and \lstinline|return|, and to support reasoning about concurrent executions, a soundness statement in ``continuation-passing'' style\footnote{Actually, continuations are not really \emph{passed} around, but rather, put onto a continuation stack, but he still calls it continuation-passing style, so we stick to this terminology here.} is preferable.
By ``continuation-passing'' style, we mean that the soundness statement roughly has the form ``\emph{for all states $\sigma_2$ satisfying condition $\phi_2$ and all continuations $k$, if it is safe to run $k$ in $\sigma_2$, then it is also safe to run $c$ followed by $k$ in any state $\sigma_1$ satisfying $\phi_1$}.''

For Hoare triples $\triple{P}{c}{Q}$, it is easy to formulate soundness in continuation-passing style by setting $\phi_1=P$ and $\phi_2=Q$.
However, it is not immediately clear how to state an information flow security property in continuation passing style, and this is the main challenge addressed in this paper.
Overall, we make the following contributions:

\begin{itemize}
\item We present a definition of information flow security in continuation passing style.
\item We state information-flow aware Hoare rules and sketch for some of them how to prove them sound with respect to our definition of information flow security.
\item We discuss how shallow embedding of the assertion language into the metalanguage combined with universal quantification can lead to a vacuous information flow security statement, and how to avoid this pitfall.
\item We present the verification of a sample C program in Coq, suggesting that verifying information flow security properties with VST-Flow will only require a minimal additional effort compared to verifying functional correctness with VST-Floyd \cite{cao2017}.
\item Finally, we discuss what it would take to reach a first sound and usable milestone of this project.
\end{itemize}



The rest of this paper is structured as follows:
Section \ref{sec:arch} explains the overall architecture and relationship to VST. It is followed by some introductory examples in section \ref{sec:examples}, and section \ref{sec:the-statement} explains how we state information flow security in continuation-passing style.
Section \ref{sec:rules} presents information flow aware Hoare rules, section \ref{sec:dimensions} discusses some design choices, and section \ref{sec:TODOs} shows what it would take to reach a first milestone.
We also discuss what could be done beyond a first milestone in section \ref{sec:future}, as well as related work in section \ref{sec:related}, and we conclude in section \ref{sec:conclusion}.

\section{Architecture}\label{sec:arch}

\tikzset{thisProj/.style={font=\bfseries}}

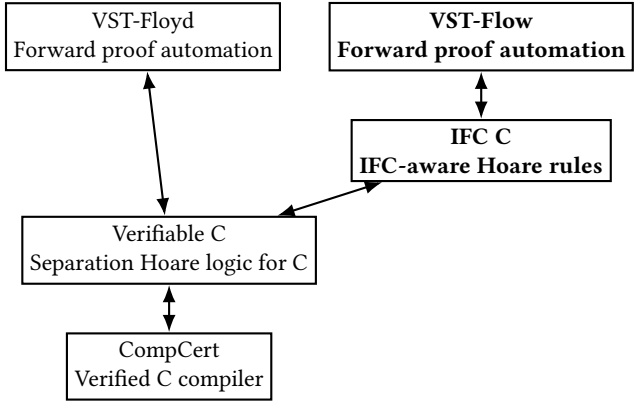
\begin{figure}
\begin{tikzpicture}[align=center, thick, node distance = 2em]
\node (veric) [draw] {Verifiable C \\ Separation Hoare logic for C };
\node (CompCert) [draw, below = of veric] {CompCert \\ Verified C compiler};
\node (ifcc) [draw, above right = of veric, thisProj] {IFC C \\ IFC-aware Hoare rules};
\node (flow) [draw, above = of ifcc, thisProj] {VST-Flow \\ Forward proof automation};
\node (floyd) [draw, left = of flow] {VST-Floyd \\ Forward proof automation};

\draw [{Latex}-{Latex}] (floyd) -- (veric);
\draw [{Latex}-{Latex}] (veric) -- (CompCert);
\draw [{Latex}-{Latex}] (flow) -- (ifcc);
\draw [{Latex}-{Latex}] (ifcc) -- (veric);
\end{tikzpicture}
\caption{Architecture (new components in bold)}\label{fig:arch}
\end{figure}

VST builds on top of CompCert, a verified C compiler which comes with formal semantics for C and assembly, and a proof that compilation preserves program behavior.
Verifiable C provides Hoare rules for C, and proves them sound with respect to CompCert's C semantics.
Since the raw Hoare rules of Verifiable C are not very convenient for users to verify programs, a library called VST-Floyd~\cite{cao2017} was built on top of this, which restates the Hoare rules in a more convenient form, and provides separation logic assertions for structs, arrays and any nesting thereof, as well as proof automation scripts to provide a simple to use verification IDE. 

Since verification of information flow properties often needs strong invariants, it is very useful to build on the functional correctness verification framework of VST.
We state information flow aware Hoare rules, which say that everything required by VST holds, as well as some additional conditions needed to guarantee information flow security.
Their soundness proof reuses the soundness proofs of Verifiable C for the functional correctness part, and only needs to prove information flow security.
On top of IFC C, we build VST-Flow, providing a similar user interface as VST-Floyd, but augmented with information flow proof support.
The proof automation part is just in the beginning of the implementation, but we expect that most parts of the VST-Floyd proof automation could be reused.

\section{Examples}\label{sec:examples}

In this paper, we will only use the two confidentiality levels $\Lo$ and $\Hi$, but our system should be extensible to any lattice with more levels.

Let us consider the following example:

\langC

\begin{lstlisting}
void f(int v, bool b, int* highptr, int* lowptr) {
  if (b) {
    *highptr = v;
  } else {
    *lowptr = v;
  }
}
\end{lstlisting}

\langCoq

The function $\cident{f}$ takes a boolean flag $\cident{b}$, which indicates whether the value $v$ is classified as $\Hi$ or $\Lo$, and depending on this, assigns it to a high-security or low-security part of the memory.

We would like to formally specify that this function does not leak $\Hi$ data into memory regions classified as $\Lo$.
In particular, this means that after running $\cident{f}$, the memory location pointed to by $\cident{*lowptr}$ should not contain any $\Hi$ data.

We will give an information flow security specification for this function.
The purpose of this example is just to give an introduction, rather than explaining all the notation details, so we will only give a minimal explanation here:

The specification is in the format $f(\overrightarrow{\texttt{y}}): \{ \trp{P} \} ~ \{ \trp{R} \} $, where $f$ is the function name, $\overrightarrow{\texttt{y}}$ are its arguments, $\trp{P}$ is a precondition consisting of a VST assertion, a stack classification (specifying the sensitivity of the data in each stack location) and a heap classification (specifying the sensitivity of the data in each heap location), and $\trp{R}$ is a postcondition in the same way.
Here's the specification:
\begin{align*}
\cident{f}(\cident{v}, \cident{b}, & \cident{highptr}, \cident{lowptr}):\\
\{\fun{x}{&((\PROP()\\
 &  \LOCAL(\ltemp{v}{x.v}, \ltemp{b}{x.b}, \ltemp{highptr}{x.h}, \ltemp{lowptr}{x.l}) \\ 
 &  \SEP(x.h \mapsto \_, x.l \mapsto \_)),\\
 &[\cident{b}: \Lo, \cident{highptr}: \Hi, \cident{lowptr}: \Lo, \cident{v}: \ite{x.b}{\Hi}{\Lo}],\\
 &[x.l: \Lo, x.h: \Hi])}\}\\ 
\{\fun{x}{&((\PROP()\LOCAL()\SEP(\\
 & x.h \mapsto (\ite{x.b}{v}{\_}), x.l \mapsto (\ite{x.b}{\_}{v}))),\\
 &[],\\
 &[x.l: \Lo, x.h: \Hi])}\}
\end{align*}

The variable $x$ is a record containing logical variables (what VST would put inside the WITH clause).
The stack classification of the precondition says that the arguments $\cident{b}$ and $\cident{lowptr}$ are classified as $\Lo$, whereas the argument $\cident{highptr}$ is classified as $\Hi$. The classification of $\cident{v}$ is \emph{value-dependent}, i.e. it depends on the value stored in the variable $\cident{b}$, which has to equal $x.b$ according to the $\LOCAL$ clause.
The heap classification says that the value stored at the location $x.l$ (that is, at the address contained in $\cident{lowptr}$) is classified as $\Lo$, and similarly, but $\Hi$, for $x.h$.
Note that we chose to keep even the address of $\cident{highptr}$ secret, but it would also be imaginable to have $[\cident{highptr}: \Lo]$ in the stack classification, but still to classify the \emph{value} stored at $\cident{highptr}$ as $\Hi$ in the heap classification.

The postcondition specifies where the value $v$ is written in its VST part, and its IFC part says that after running the function, $x.l$ still only contains $\Lo$ data.

Such functions are typical use cases of the Cross Domain Desktop Compositor \cite{beaumont2016}, a screen compositor connected to two machines of different security levels, allowing to display data from these two machines on the same screen and guaranteeing that no data from the confidential machine leaks into the less confidential one.

TODO example inspired by CDDC

It should be an example ``where abstraction makes no sense'', i.e. where reasoning at the low-level is the most natural thing to do, where the security property is tied to the physical representation of the data.


\section{The information flow security statement}\label{sec:the-statement}

This section explains our information flow judgment, starting from the classical Hoare triple, then explaining the additions made by VST and then our additions to reason about information flow security.

\subsection{Hoare logic}

Hoare logic \cite{hoare1969} uses triples of the form
\[ \triple{P}{c}{R} \]
intended to mean that if we are in a state where $P$ holds and run the command $c$, we will end up in a state where $R$ holds.

\subsection{VST logic}

VST adds a context $\Delta$ to this judgment, where $\Delta$ gives the types of function parameters, local and global variables, and specifications of global functions.


We will use the type alias \lstinline|vst_assert| for the type of VST assertions.
Conceptually, they can be thought of as a function taking a stack and a heap and returning a proposition, but the actual type is a bit more complex because of VST's step indexing.

While VST does not support \lstinline|goto| commands, it does support \lstinline|break|, \lstinline|continue| and \lstinline|return| before the end of the function body, and to account for these premature ways of exiting a block of code, VST's postconditions are not just of type \lstinline|vst_assert|, but of type \lstinline|exitkind->option val->vst_assert|, where the type \lstinline|exitkind| is an enum with the four values $\EKnormal$, $\EKbreak$, $\EKcontinue$, and $\EKreturn$ (to denote normal code execution until the end of the code block, or premature exit via \lstinline|break|, \lstinline|continue| or \lstinline|return|, respectively), and the \lstinline|option val| is used for the return value if there is one.
Alternatively, VST postconditions can be thought of as a quadruple of four assertions, one for each possible way of exiting a code block.

So, VST judgments are written
\[ \vsttriple{\Delta}{P}{c}{R} \]
where $P$ is of type \lstinline|vst_assert|, $c$ is a C statement, and $R$ is of type \lstinline|exitkind->option val->vst_assert|.

We will use $\lambda$ notation to construct functions, so a VST judgment might look like
\[ \vsttriple{\Delta}{P}{c}{\fun{ek}{\fun{v}{P'}}} \]
where $ek$ is an argument of type \lstinline|exitkind| and $v$ is an argument of type \lstinline|option val|.

In order to give useful specifications, one usually has to connect values of the precondition with values of the postcondition.
Since VST's assertion language is shallowly embedded into the metalanguage (i.e. into Coq), we can use universal quantification of the metalanguage to achieve this, so VST judgments actually have the form
\[ \forall x:T.~ \vsttriple{\Delta}{\ap{P}{x}}{c}{\ap{R}{x}} \]
where $T$ is a type that the user who writes the specification picks; typically it is a tuple type, because several metavariables are needed to connect the pre- and postcondition.

\subsection{Adding information flow control}

To add support for reasoning about information flow security on top of this, we need the following definitions:

The type \lstinline|label| has the two members \lstinline|Lo| and \lstinline|Hi|, used to label data of low and high security, respectively.
Our goal will be to prove that data labeled as \lstinline|Hi| does not influence the values of data labeled as \lstinline|Lo|.

Stack classification functions take the name of a stack variable and return a \lstinline|label| for it. We will use the letter $N$ (for non-addressable) to denote stack classification functions, and we introduce the type \lstinline|stack_clsf| as an alias for the function type \lstinline|ident->label|.

We introduce the type \lstinline|heap_loc| for heap locations.
Heap classification functions take such a \lstinline|heap_loc| and return a \lstinline|label| for it. We will use the letter $A$ (for addressable) to denote heap classification functions, and we introduce the type \lstinline|heap_clsf| as an alias for the function type \lstinline|heap_loc->label|.

We now can augment VST assertions to information flow control (IFC) assertions, by defining the type \lstinline|ifc_assert| as an alias for triples of a \lstinline|vst_assert|, a \lstinline|stack_clsf|, and a \lstinline|heap_clsf|, so information flow control judgments will have the form
\[ \ifcdef{\Delta}{P,N,A}{c}{\fun{ek}{\fun{v}{(P', N', A')}}} \]
We will use calligraphic letters $\trp{P}$ and $\trp{Q}$ for IFC preconditions, and $\trp{R}$ for IFC postcondition functions depending on an \lstinline|exitkind| and a return value, so whenever we do not need to access the three parts of IFC assertions separately, we will write judgments of the form
\[ \ifcdef{\Delta}{\trp{P}}{c}{\trp{R}} \]

Notice that there is an interesting difference between VST's assertion $P$ and the classification functions $N$ and $A$:
VST's $P$ carries its meaning in its definition, that is, if $\sigma$ is a program state, then $\ap{P}{\sigma}$ is a proposition (in the metalanguage) expressing some requirements on the values of $\sigma$.
On the other hand, $N$ and $A$ are purely syntactic constructs whose meaning will only be defined by the semantic definition of the IFC judgment.
That is, $P$ is shallowly embedded, while $N$ and $A$ are deeply embedded.
The latter is not a design choice, because $N$ and $A$ can only be given meaning in a bigger context where a pre- and postcondition is available.

\subsection{C light operational semantics}

We will define information flow security with respect to VST's small-step operational semantics for C light.
States $\sigma$ are triples $\clstate{e}{k}{m}$ of a variable environment $e$ representing the values on the stack, a continuation stack $k$ consisting of the commands to execute, and a heap memory $m$.

We write $\clstep{\sigma_1}{\sigma_2}$ for the small-step reduction relation,
and we write $\clstar{\sigma_1}{\sigma_2}$ for its transitive closure,
and $\clstarN{\sigma_1}{n}{\sigma_2}$ to say that after $n$ steps, state $\sigma_1$ ends up in state $\sigma_2$.
Moreover, we define execution until final state, written $\clbigstep{\sigma_1}{\sigma_2}$, as $\clstar{\sigma_1}{\sigma_2}$ where the command to be executed in $\sigma_2$ is the empty command, which means that execution is done (and hasn't got stuck along the way).

\subsection{Defining the semantics of the IFC judgment}

In the subsequent sections, we will define the meaning of the IFC judgment iteratively, by first presenting some simple definitions which look promising, but turn out to be dead ends.
This will then justify our final, slightly more complex definition.

\subsection{Semantics of IFC judgment: First attempt}

In order to define the meaning of the IFC judgment, we need the following auxiliary definition:
\begin{definition}[Simple low-equivalence]\label{def:simple-lo-equiv}
Two states $\clstate{e}{k}{m}$ and $\clstate{e'}{k'}{m'}$ are called low-equivalent with respect to the classification functions $N$ and $A$ (of type \lstinline|stack_clsf| and \lstinline|heap_clsf|, respectively) if for all stack locations $\ell$ for which $N(\ell)=\Lo$, $e(\ell)=e'(\ell)$ and for all heap locations $\ell$ for which $A(\ell)=\Lo$, $m(\ell)=m'(\ell)$.
\end{definition}


We will treat postconditions as if they were of type \lstinline|ifc_assert| (instead of \lstinline|exitkind->option val->ifc_assert|) for this first definition attempt, so we ignore premature exits, because the problem we want to illustrate here already appears in this simpler setting.

\begin{definition}[Meaning of IFC judgment, first attempt]\label{def:first-attempt}
The meaning of $\ifcdef{\Delta}{P_1,N_1,A_1}{c}{P_2,N_2,A_2}$ is: The VST judgment $\vsttriple{\Delta}{P_1}{c}{P_2}$ holds and for all $\sigma_1$, $\sigma_1'$, $\sigma_2$, $\sigma_2'$, if $\ap{P_1}{\sigma_1}$ and $\ap{P_1}{\sigma_1'}$ hold and both in $\sigma_1$ and $\sigma_1'$, the command to be executed is $c$, 
and $\sigma_1$ is low-equivalent to $\sigma_1'$ w.r.t $N_1$ and $A_1$,
and $\clbigstep{\sigma_1}{\sigma_2}$ and $\clbigstep{\sigma_1'}{\sigma_2'}$, then 
$\sigma_2$ is low-equivalent to $\sigma_2'$ w.r.t $N_2$ and $A_2$.
\end{definition}

Proving this statement for a particular program $c$ would then prove information flow security for that program in the sense that $\Hi$ data does not influence the values of $\Lo$ data, because if we vary the values of $\Hi$ data between $\sigma_1$ and $\sigma_1'$, these changes can only result in variations of $\Hi$ data between $\sigma_2$ and $\sigma_2'$, but cannot influence $\Lo$ values of $\sigma_2$ and $\sigma_2'$, because they have to be equal.

\subsection{The problem with universal quantification over an IFC judgment}

If we use universal quantification to connect values of the precondition with values of the postcondition, as commonly done in VST, we don't get the statement that we want.
Let's consider the following example:
\begin{align*}
\forall x:\cident{int}.~ \Delta~ & \{\cident{sec} = x, [\cident{sec}: \Hi, \cident{pub}: \Lo ], A\}\\
 & \texttt{pub := sec} \\
 & \{\cident{sec} = x \wedge \cident{pub} = x, [\cident{sec}: \Hi, \cident{pub}: \Lo ], A\}
\end{align*}
where we use square brackets to define the stack classification functions by enumerating their values for all variable names, and $A$ is an unspecified irrelevant heap classification function.

Clearly, this statement should not be provable, because it assigns the secret $\Hi$ variable $\cident{sec}$ to the public $\Lo$ variable $\cident{pub}$.
However, it \emph{is} provable, because to prove the universal quantification, we assume $x$ to be an arbitrary, but \emph{fixed} value, so the $\Hi$ variable $\cident{sec}$ cannot have different values in the states $\sigma_1$ and $\sigma_1'$ from Definition \ref{def:first-attempt}, and therefore, $\cident{pub}$ will always have the same value in $\sigma_2$ and $\sigma_2'$, which makes the statement true.

So we see that the way we combined universal quantification with our definition of information flow security is flawed, because it does not express what we want it to express.

\subsection{Adapting the shape of the IFC judgment}

Therefore, we have to give control over the quantification to the IFC judgment, rather than adding it on the outside.
To do so, we parameterize all pre- and postconditions by a variable $x$ of type $T$, which can be any user-specified tuple type.
That is, preconditions now have the type \lstinline|T->ifc_assert|, and postconditions have the type \lstinline|T->exitkind->option val->ifc_assert|, and IFC judgments will typically look like
\[ \ifcdef{\Delta}{\fun{x}{(P,N,A)}}{c}{\fun{x}{\fun{ek}{\fun{v}{(P',N',A')}}}} \]
Note that $P$ links $x$ to values on the stack and heap, so allowing the classification functions to depend on $x$ allows for the same kind of value-dependent classifications as Murray et al \citep{murray2016} and Costanzo and Shao \cite{costanzo2014} have. For instance, the classification of one variable might depend on a boolean flag stored in another variable.

This change now allows the IFC judgment to ``see'' the dependency of the pre- and postcondition on $x$, so its definition will be able to vary the value of the $x$ it passes to them in the same way it varies the state $\sigma_1$ to $\sigma_1'$.
This will allow variations of the values of $\Hi$ variables between $\sigma_1$ to $\sigma_1'$ without the precondition $P$ preventing them, because we require that $\app{P}{x}{\sigma_1}$ and $\app{P}{x'}{\sigma_1'}$ hold, instead of requiring that the same $P$ holds for both $\sigma_1$ and $\sigma_1'$ as we did in Definition \ref{def:first-attempt}.

\subsection{Semantics of IFC judgment: Second attempt}

Now, since $N$ and $A$ depend on an $x$ that we want to vary to an $x'$, we might get two different classifications: One variable might be classified as $\Lo$ by $\ap{N}{x}$ but as $\Hi$ by $\ap{N}{x'}$. So we adapt the definition of low-equivalence as follows:

\begin{definition}[Low-equivalence]
Two states $\sigma = \clstate{e}{k}{m}$ and $\sigma' = \clstate{e'}{k'}{m'}$ are called low-equivalent with respect to the classification functions $N$, $N'$ and $A$, $A'$ if for all stack locations $\ell$ for which $\ap{N}{\ell}=\Lo=\ap{N'}{\ell}$, we have $\ap{e}{\ell}=\ap{e'}{\ell}$ and for all heap locations $\ell$ for which $\ap{A}{\ell}=\Lo=\ap{A'}{\ell}$, we have $\ap{m}{\ell}=\ap{m'}{\ell}$.
\end{definition}


Note that we only require equality if \emph{both} classifications are $\Lo$.
Costanzo and Shao \cite{costanzo2014} do so as well, and it seems that this is the most useful definition: It is strong enough to obtain a meaningful statement, because one just has to make sure that the locations we care about have constant $\Lo$ classifications.

As in the first attempt, we temporarily ignore premature exits in the following definition:

\begin{definition}[Meaning of IFC judgment, second attempt]\label{def:attempt-x-no-cont}
The meaning of $\ifcdef{\Delta}{\fun{x}{(P_1,N_1,A_1)}}{c}{\fun{x}{(P_2,N_2,A_2)}}$ is: For all $x$, the VST judgment $\vsttriple{\Delta}{\ap{P_1}{x}}{c}{\ap{P_2}{x}}$ holds and for all $x$, $x'$, $\sigma_1$, $\sigma_1'$, $\sigma_2$, $\sigma_2'$, if $\app{P_1}{x}{\sigma_1}$ and $\app{P_1}{x'}{\sigma_1'}$ hold and both in $\sigma_1$ and $\sigma_1'$, the command to be executed is $c$, 
and $\sigma_1$ is low-equivalent to $\sigma_1'$ w.r.t $\ap{N_1}{x}$, $\ap{N_1}{x'}$ and $\ap{A_1}{x}$, $\ap{A_1}{x'}$
and $\clbigstep{\sigma_1}{\sigma_2}$ and $\clbigstep{\sigma_1'}{\sigma_2'}$, then 
$\sigma_2$ is low-equivalent to $\sigma_2'$ w.r.t $\ap{N_2}{x}$, $\ap{N_2}{x'}$ and $\ap{A_2}{x}$, $\ap{A_2}{x'}$.
\end{definition}

As one can see, the VST part of the definition only quantifies over one $x$, whereas the IFC part of the definition quantifies over two different $x$ and $x'$.

This definition might be fine, but it does not account for premature exits, so we have to look at them in the following subsections.

\subsection{C light operational semantics for break, continue, and return}

VST's C light operational semantics include a stack of continuations to be executed in the program state.
A continuation can be a sequence of two statements to be resumed at the second statement, a loop to be resumed at its increment statement, a loop to be resumed at its body, or a function body to be resumed after a return.

That is, a program state $\sigma$ can be written as $\clstate{e}{k}{m}$, where $e$ is the environment of local stack variables, $k$ is the continuation stack, and $m$ is the heap memory.\footnote{In fact, VST also defines a \lstinline|corestate| for the case where execution is in an external functions, but we ignore this for the moment.}

VST defines a function \lstinline|exit_cont|, which takes an \lstinline|exitkind|, an optional return value, and a continuation stack, and pops the continuation stack according to the \lstinline|exitkind|, pushing the return value into the environment of the calling function in the case of $\EKreturn$.

Instead of presenting all the details of VST's operational semantics, we just show one example, the rule for $\cbreak$:
\[\inference[\textsc{step-break}]{
  \clstep{\clstate{e}{\mbox{\lstinline|exit_cont|}~\EKbreak~\mbox{\lstinline|None|}~k}{m}}{\sigma}
}{
  \clstep{\clstate{e}{\cbreak::k}{m}}{\sigma}
}\]
The double colon stands for list cons of the continuation stack, and \lstinline|(exit_cont $\EKbreak$ None $k$)| is defined to pop the continuation stack $k$ until it encounters a continuation standing for a loop to be resumed at its increment statement, which it also pops, so that execution will continue after the loop.

\subsection{Third attempt: Adding premature exits}

Using the \lstinline|exit_cont| function, we can now add support for premature exits to our information flow security definition:

\begin{definition}[Meaning of IFC judgment, third attempt]\label{def:attempt-x-ek-no-cont}
The meaning of $\ifcdef{\Delta}{\fun{x}{(P_1,N_1,A_1)}}{c}{\fun{x}{\fun{ek}{\fun{v}{(P_2,N_2,A_2)}}}}$ is: For all $x$, the VST judgment $\vsttriple{\Delta}{\ap{P_1}{x}}{c}{\ap{P_2}{x}}$ holds and for all $x, x', ek, v, v', e_1, e_1', m_1, m_1', k, e_2, e_2', m_2, m_2'$, if $\app{P_1}{x}{\clstate{e_1}{c::k}{m_1}}$ and $\app{P_1}{x'}{\clstate{e_1'}{c::k}{m_1'}}$ hold,
and $e_1$ is low-equivalent to $e_1'$ w.r.t. $\ap{N_1}{x}$, $\ap{N_1}{x'}$,
and $m_1$ is low-equivalent to $m_1'$ w.r.t. $\ap{A_1}{x}$, $\ap{A_1}{x'}$,
and $\clstar{\clstate{e_1}{c::k}{m_1}}{\clstate{e_2}{\appp{\excont}{ek}{v}{k}}{m_2}}$
and $\clstar{\clstate{e_1'}{c::k}{m_1'}}{\clstate{e_2'}{\appp{\excont}{ek}{v'}{k}}{m_2'}}$
then $e_2$ is low-equivalent to $e_2'$ w.r.t. $\ap{N_2}{x}$, $\ap{N_2}{x'}$,
and $m_2$ is low-equivalent to $m_2'$ w.r.t. $\ap{A_2}{x}$, $\ap{A_2}{x'}$.
\end{definition}

Note that we now have to consider an execution from a continuation stack of the form $c::k$ to the stack $\appp{\excont}{ek}{v}{k}$, rather than from a stack containing only the command $c$ to an empty stack, because eg. in $\ifcdef{\Delta}{\trp{P}}{\cbreak; c_2}{\trp{R}}$, the $\cbreak$ might skip more commands than just $c_2$.

\subsection{Problems inverting multistep}\label{sec:cannot-invert-multistep}

Trying to prove Hoare rules sound with respect to Definition \ref{def:attempt-x-ek-no-cont} revealed the following problem:
Such proofs have to invert hypotheses of the form $\clstar{\clstate{e_1}{c::k}{m_1}}{\clstate{e_2}{\appp{\excont}{ek}{v}{k}}{m_2}}$ to get information about how the command $c$ was executed.
This is problematic even in the simplest case where $ek=\EKnormal$, which simplifies to $\clstar{\clstate{e_1}{c::k}{m_1}}{\clstate{e_2}{k}{m_2}}$:
If $c$ is inside a loop body, and $k$ contains a loop continuation, it might have happened that after running $c$ and arriving at continuation stack $k$, another loop iteration was performed, going through a state with continuation stack $c::k$ again, and then arriving at continuation stack $k$.
That is, the commands executed during $\clstar{\clstate{e_1}{c::k}{m_1}}{\clstate{e_2}{k}{m_2}}$ could be not just $c$, but also parts of $k$, but all available hypotheses only talk about the execution of $c$, not $k$.

So $\clstar{\clstate{e_1}{c::k}{m_1}}{\clstate{e_2}{k}{m_2}}$ does not express what we want it to express, and we have to find a better definition for the meaning of the IFC judgment.

\subsection{VST's guard-based soundness proof}

Similar problems must have occurred in the VST soundness proof, and VST solves them as follows:
Contrary to what one might expect, VST does \emph{not} define the meaning of its Hoare judgment as follows:

\begin{definition}[Meaning of Hoare judgment, unusable definition]
The meaning of $\vsttriple{\Delta}{P}{c}{R}$ is:
For all $e_1, m_1, k$,
if $\ap{P}{\clstate{e_1}{c::k}{m_1}}$ holds,
then there exist $ek, v, e_2, m_2$ such that
$\clstar{\clstate{e_1}{c::k}{m_1}}{\clstate{e_2}{\appp{\excont}{ek}{v}{k}}{m_2}}$
and $\appp{R}{ek}{v}{\clstate{e_2}{\appp{\excont}{ek}{v}{k}}{m_2}}$ holds.
\end{definition}

Instead, it is defined with the following series of definitions:

\begin{definition}[Immediately safe]
A state $\sigma=\clstate{e}{k}{m}$ is immediately safe if $k=\nil$ or $\clstep{\sigma}{\sigma_2}$ for some $\sigma_2$.
\end{definition}

\begin{definition}[Safe]
A state $\sigma$ is safe if for all $\sigma_2$, if $\clstar{\sigma}{\sigma_2}$, then $\sigma_2$ is immediately safe.
\end{definition}

\begin{definition}[Guard]
Predicate $P$ guards the continuation stack $k$, written $\guard{P}{k}$, if for all $e$ and $m$, $\ap{P}{\clstate{e}{k}{m}}$ implies that $\clstate{e}{k}{m}$ is safe.
\end{definition}

\begin{definition}[Return guard]
Postcondition $R$ guards the continuation stack $k$, written $\rguard{R}{k}$, if for all $ek$ and $v$, we have $\guard{\app{R}{ek}{v}}{k}$.
\end{definition}

\begin{definition}[Meaning of Hoare judgment]
The meaning of $\vsttriple{\Delta}{P}{c}{R}$ is:
For all $k$, $\rguard{R}{k}$ implies $\guard{P}{(c::k)}$.
\end{definition}

It might look like the above definition only talks about safety in the sense of absence of crashes, but in fact, it does guarantee functional correctness, because $k$ could be any program which tests whether $R$ holds, and crashes if it does not hold.
Then, the above definition guarantees that after running $c$, $R$ must hold.


\subsection{Final definition of the IFC judgment}

\begin{definition}[Equivalent continuations]
Two continuations $c_1$ and $c_2$ are called equivalent, written $\contequiv{c_1}{c_2}$, if they are equal or they are both a function body to be resumed after a return, of the same function, but with potentially different variable environments to be restored.
\end{definition}

\begin{definition}[Head-equivalent states]
Two states $\sigma = \clstate{e}{k}{m}$ and $\sigma' = \clstate{e'}{k'}{m'}$ are called head-equivalent, written
$\cscontheadequiv{\sigma}{\sigma'}$ if either both $k$ and $k'$ are the empty stack, or both are non-empty and their head (top) continuations are equivalent.
\end{definition}

\begin{definition}[Sync]\label{def:sync}
Two states $\sigma_1$ and $\sigma_1'$ are called ``in sync'',\footnote{We are still looking for a better name for this definition, suggestions are welcome!} written $\sync{\sigma_1}{\sigma_1'}$, if for all $n, \sigma_2, \sigma_2'$, if $\clstarN{\sigma_1}{n}{\sigma_2}$ and $\clstarN{\sigma_1'}{n}{\sigma_2'}$, then $\cscontheadequiv{\sigma_2}{\sigma_2'}$.
\end{definition}

Sync can be thought of as some kind of low-equivalence, with the advantage that it does not need any classification functions, which are typically only available for the program state right before and right after the command in question, but not for intermediate states or future states.

In fact, low-equivalence between two memories for a bit stored at heap location $\ell$ can be encoded as follows using sync:
Let $k$ be a continuation stack whose program loads the bit at location $\ell$ and then branches on the value of that bit, executing some command $c_0$ if it is 0, or some different command $c_1$ (such that $\contequiv{c_0}{c_1}$ does \emph{not} hold) if it is 1.
Now if we have two variable environments $e_1$ and $e_1'$, and two memories $m_1$ and $m_1'$, and we want to say that after running some given command $c$, the bit at $\ell$ must be the same in both memories, we can express this as $\sync{\clstate{e_1}{c::k}{m_1}}{\clstate{e_1'}{c::k}{m_1'}}$.
If $c$ terminates, it does so in a certain number of steps $n$, and after $n+1$ steps, execution will be in $k$ and branch on the value stored at $\ell$, putting $c_0$ or $c_1$ on top of the continuation stack depending on the bit stored at $\ell$, and since sync requires the two continuation stack heads to be equivalent, it is ensured that the value stored at $\ell$ in the two memories is the same.

That is, we can append a ``test continuation'' $k$ to the command $c$ in question, which makes the sync proposition false if any equality we desire to hold does not hold.

We can use this intuition to define an IFC guard in a similar way as VST's guard:


\begin{definition}[IFC guard]
We write $\iguard{\fun{x}{(P,N,A)}}{k}{k'}$ if for all $x, x', e, e', m, m'$, if $\app{P}{x}{\clstate{e}{k}{m}}$ and $\app{P}{x'}{\clstate{e'}{k'}{m'}}$ hold, and $e$ is low-equivalent to $e'$ w.r.t. $\ap{N}{x},\ap{N}{x'}$ and $m$ is low-equivalent to $m'$ w.r.t. $\ap{A}{x},\ap{A}{x'}$, then $\sync{\clstate{e}{k}{m}}{\clstate{e'}{k'}{m'}}$.
\end{definition}

\begin{definition}[IFC return guard]
We write $\irguard{\trp{R}}{k}{k'}$ if for all $ek$ and $v$, we have $\iguard{\fun{x}{\appp{\trp{R}}{x}{ek}{v}}}{k}{k'}$.
\end{definition}

\begin{definition}[Meaning of IFC judgment, final version]
The meaning of $\ifcdef{\Delta}{\trp{P}}{c}{\trp{R}}$ is: 
For all $x$, the VST judgment $\vsttriple{\Delta}{\ap{P_1}{x}}{c}{\ap{P_2}{x}}$ holds
and for all $k$ and $k'$, $\irguard{\trp{R}}{k}{k'}$ implies $\iguard{\trp{P}}{(c::k)}{(c::k')}$.
\end{definition}

Note: VST has one forall $x$ in front of the whole judgment, whereas IFC has a forall inside both guards (the iguard and the irguard).

\subsection{Discussion}

This definition crucially depends on the restriction that branching on $\Hi$ data is not allowed.

\section{Rules}\label{sec:rules}

\subsection{The lattice}

To classify the security level of values, we use the two-point lattice called \lstinline|label|, whose element $\Lo$ is the bottom element $\bot$ and $\Hi$ is the top element $\top$.
We write $\lle{\Lo}{\Hi}$ for the lattice order, $\lub{\ell_1}{\ell_2}$ for the least upper bound, and $\glb{\ell_1}{\ell_2}$ for the greatest lower bound.

If $L$ is a lattice, then for any type $T$, the type $T \rightarrow L$ is a lattice as well, and we use the same operators for lattices lifted over tuple types, exitkinds, stack locations and heap locations.

\subsection{Expression classification}

We define the function $\mathcoq{clsf\_expr}$, which takes a stack classification $N$ and a VST C light expression $e$ and returns a label giving the highest classification of any variable occurring in $e$.
Since expressions are pure, heap-independent, and cannot make function calls, it suffices to give $N$ as an argument, and no heap classification is needed.
Together with $\mathcoq{clsf\_expr}$, we also define $\mathcoq{clsf\_lvalue}$, which classifies an l-value expression, and finally, we define $\mathcoq{clsf\_exprs}$ to work on a list of expressions.

\begin{figure}
\input{ifc-loop.tex}
\input{ifc-seq.tex}
\input{ifc-break.tex}
\input{ifc-continue.tex}
\input{ifc-ifthenelse.tex}
\caption{Control flow rules (with explicit logical variable $x$)}\label{fig:control-flow-rules}
\end{figure}

\subsection{Notational conventions}

The enum type \lstinline|exitkind| has the four values $\EKnormal$, $\EKbreak$, $\EKcontinue$, and $\EKreturn$.
Given a variable $ek$ holding an exit kind, we define the notation $(f_1, f_2, f_3, f_4)_{ek}$ to mean

\begin{center}
\begin{tabular}{c}
\begin{lstlisting}
match $ek$ with
| $\EKnormal$ => $f_1$
| $\EKbreak$ => $f_2$
| $\EKcontinue$ => $f_3$
| $\EKreturn$ => $f_4$
end
\end{lstlisting}
\end{tabular}
\end{center}

in order to select the appropriate postcondition from a quadruple with a postcondition for each exitkind.


We define the operator $\mathcoq{nret}$ to assert that a command exits normally:
\[\nret{\trp{P}} := \fun{ek}{\fun{v}{\ite{ek=\EKnormal}{\trp{P}}{(\bot, \bot, \bot)}}}\]
Note that the first $\bot$ means false (a VST assertion), the second $\bot$ is an element of the lifted lattice \lstinline|T->ident->label|, and the third bot is an element of the lifted lattice \lstinline|T->heap_loc->label|.


We use the notation $f[x := v]$ for function update, i.e.
\[f[x := v] := \fun{x_0}{\ite{x_0=x}{v}{\ap{f}{x_0}}} \]
and in examples, we use square brackets for function literals, e.g.
\[ [\cident{sec}: \Hi, \cident{pub}: \Lo ] \]
To make such functions total, we assume that they return the default value $\Hi$ for undefined locations.

\subsection{The rules}

The control flow rules are given in Figure \ref{fig:control-flow-rules}.
They are the same in VST, except that all assertions also abstract over a logical variable $x$ of some user-defined type $T$.
In \textsc{ifc-if}, we require that the expression $b$ does not contain any $\Hi$ values, to make sure that the low-equivalent states do not take different execution paths.
Note that to prove this, one may assume that the precondition $\ap{P}{x}$ holds.
This is important in the case of value-dependent classifications, where the classification might only be low if some requirements on $x$ (encoded by $P$) hold.

To keep the notation sane in Figures \ref{fig:consequence-rules} and \ref{fig:straight-rules}, we assume implicit lifting of all operators over an argument $x:T$, where $T$ is a user-defined record of logical variables.
For instance, the $\vdash$ operator refers now to the one from lifted separation logic, so $P \vdash Q$ really means $\forall x:T.\, \ap{P}{x} \vdash \ap{Q}{x}$, and the triple constructor is assumed to be lifted, that is, $\{P,N,A\}$ stands for $\{\fun{x}{(\ap{P}{x},\ap{N}{x},\ap{A}{x}))}\}$.

\begin{figure}
\input{ifc-pre.tex}
\input{ifc-post}
\caption{Consequence rules (with implicit lifting over $x$)}\label{fig:consequence-rules}
\end{figure}

\begin{figure}
\input{ifc-set.tex}
\input{ifc-load.tex}
\input{ifc-store.tex}
\caption{Rules for assignment, load and store (with implicit lifting over $x$)}\label{fig:straight-rules}
\end{figure}

\begin{figure}
\input{ifc-call.tex}
\input{ifc-return.tex}
\caption{Rules for function call and return (WIP)}\label{fig:call-rules}
\end{figure}

\subsection{Function specifications}

Function specifications are a quintuple $((f,s),N,A,N',A')$ (or a record in Coq) consisting of a VST function specification $s$ and a function name $f$, and pre- and post-classification functions for stack and heap. Note that the only purpose of $N'$ is to classify the return value, the other local variables don't matter after the return.
They all are parameterized by a logical variable $x$ of type $T$, where $T$ is the same type as the WITH-clause type of $f$.

Instead of writing function specifications as such a quintuple, we will write
\[ f(\overrightarrow{\texttt{y}}): \{\funtpd{x}{T}{(P,N,A)}\} ~ \{\funtpd{x}{T}{(P',N',A')}\}  \]
Note that contrary to VST, we do not use a $\Pi$ to quantify over an $x$ of type $T$, because that's left to the definition of the meaning of the IFC judgment.


\subsection{Call rules}

Note that the call rule in Figure \ref{fig:call-rules} is work in progress.
We expect that $\Hi$ could be used as a default for classifying heap parts outside the heap accessible to the function, and then the greatest lower bound $\glb{A_0}{A}$ could be used to ``join''\footnote{Unfortunately, the pun doesn't work, it's just the wrong way round: Greatest lower bound is a synonym for meet, not join.} two heap classifications.

Also, we should investigate how the logical $(x:T_0)$ from caller connects to the logical $(x:T)$ from the callee.

\subsection{Rules in canonical form}

Some rules in canonical form were developed in Coq, one of which is shown in Figure \ref{fig:canon-rules}.

\begin{figure}
\input{ifc-load-canon.tex}
\caption{Canonical load rule}\label{fig:canon-rules}
\end{figure}

\section{Design considerations/dimensions}\label{sec:dimensions}

\subsection{Modeling premature exits}

No matter whether the operational semantics are small-step or big-step, there are two ways to model premature exits such as \lstinline|break|, \lstinline|continue|, \lstinline|return| before the end of the function body, or exceptions:

\paragraph{With immediately jumping exit commands}
This is the approach used in VST: For instance, the \lstinline|break| command jumps directly outside the loop (without even ``consuming'' an execution step), and similarly for the other exit commands.

\paragraph{With extended states storing the exitkind}
In this approach, one defines an \emph{extended state} to be either a \emph{normal} state, or an \emph{abrupt} state. Premature exit commands step from a normal state to an abrupt state, and only the next step specifies how to continue after the the abrupt state.
This approach is used in seL4. It is described in Norbert Schirmer's PhD thesis \cite{schirmer2006} for both small-step and big-step semantics.

\vspace{1em}

As explained in section \ref{sec:cannot-invert-multistep}, the VST approach has the problem that neither $\clstar{\clstate{e_1}{c::k}{m_1}}{\clstate{e_2}{k}{m_2}}$ nor $\clstar{\clstate{e_1}{c}{m_1}}{\clstate{e_2}{\nil}{m_2}}$ expresses what we need for a direct-style definition of the meaning of a Hoare triple.
The seL4 approach, however, does not suffer from this problem, because the hypothesis $(\textit{Normal}~s, c) \rightarrow^\ast (\textit{Abrupt}~s', \nil)$ exactly captures what we want to say, because it can store an exitkind inside the state.

\subsection{Direct-style vs guard-style}

We decided to use a guard-style definition (as in Appel and Blazy~\cite{appel2007} and used in VST) for our IFC Hoare judgment, because of the reasons outlined in section \ref{sec:the-statement} and for better compatibility with VST.

However, a direct-style definition would be more understandable for users.

It should be investigated further what implications this design decision has on proofs about concurrent programs, and whether it is possible to write proofs that the guard-style definition implies a direct-style definition.

\subsection{Assertions about intermediate states}

If we use a small-step direct-style definition, the question arises whether the definition of the IFC Hoare judgment should make an assertion only about the final state reached after executing the statement in question, or also about all intermediate steps.
This might have implications on proofs, especially proofs about compositionality for concurrency.

However, we have to keep in mind that the IFC Hoare judgment is only provided classification functions for the initial and final state, but not for the intermediate states, so if some kind of low-equivalence is to be asserted about intermediate states, this cannot use classification functions.
One way to deal with this problem is to use a judgment like \emph{sync} (Definition \ref{def:sync}), which makes some guarantees without needing classification functions.

\subsection{How to define what we prove}

We can distinguish two ways to give the semantics of non-interference proofs:

\paragraph{With proof rules proven sound}
In this approach one defines a semax judgment, where the right-hand side of the definition is the non-interference property we want to prove.
Then, one proves several rules, typically one for each kind of statement.
This approach is used in this paper, by VST \cite{VST} and by the seL4 non-interference proofs \cite{murray2013}.

\paragraph{With typing derivations}
In this approach, one defines a typing judgment using inference rules, and writes a soundness proof stating that if a typing derivation for a program can be found, then the desired non-interference property holds. This approach is used in Murray's CSF 16 paper \cite{murray2016}, and also by Costanzo and Shao \cite{costanzo2014}.

\section{Towards a first milestone}\label{sec:TODOs}

\subsection{Next steps}


\paragraph{Investigate function calls} How does the stack classification change before and after function calls? \textsc{ifc-return} does not need to deal with this, because it only goes to an equivalent state which has not returned yet, so \textsc{ifc-call} does all the work here, and this should work, because the rule ``knows'' both the ``outer'' and ``inner'' stack. There is an \textsc{ifc-call} rule on paper, but it should be implemented in Coq and tried out with an example.

\paragraph{Guard-style vs direct-style} Does guard-style soundness imply direct-style soundness for VST? For IFC? Can we prove this in Coq?

\paragraph{Finish proving IFC Hoare rules} Proving soundness of the IFC Hoare rules w.r.t. our IFC definition.

\paragraph{Step-indexed assertions} Currently, all the IFC development is using assertions of type \lstinline|env->temp_env->mem->Prop| instead of \lstinline|environ->mpred|, and admits a conversion function between the two, as well as some properties of this conversion function, none of which probably holds. This gap should be closed, that is, we should use \lstinline|environ->mpred| in the IFC development as well.

\paragraph{Connect basic IFC rules with IFC rules in canonical form} Currently, some basic IFC rules are used directly in the examples, while for others, an IFC rule in canonical form is admitted. For all commands, we should have IFC rules in canonical form, and prove them w.r.t. to the basic IFC rules.
It will be interesting to see how/whether parts of the Floyd proofs can be reused for this.

\paragraph{More examples} We should verify more and larger examples to evaluate the usability and expressivity of the system.

\subsection{Pending implementation tasks}

\paragraph{Lifting over logical variables} All assertions need to be lifted over a record containing logical variables. Some investigations in the file \lstinline|ifc/proofauto_lemmas.v| suggest that the same lifting infrastructure that VST uses to lift assertions over the \lstinline|environ| could be used to additionally lift the assertions over a record of logical variables, but this has not yet been applied to the whole development.

\paragraph{Include IFC information in $\Delta$} The environment $\Delta$ should now contain \lstinline|ifc_funspec|s instead of VST \lstinline|funspec|s. It will be interesting to see how well the existing VST-Floyd proof automation infrastructure supports this change.

\section{Future work}\label{sec:future}

\paragraph{Influence of VST on our definitions} It would be interesting to ask what our definitions would look like if we were starting from scratch, i.e. not attempting to reuse the VST soundness proof. Does that statement follow from the one we're proposing?

\paragraph{Declassification} Can we support declassification by preconditions in the same neat way as Costanzo and Shao \cite{costanzo2014}? Probably not directly, because they need to attach labels to logical variables to make it work, which we currently don't do.

\paragraph{Branching on $\Hi$ data} Is a two-execution semantics needed for this, bisimulations?  Or can we get away with just a better definition of \emph{sync}? Instead of asserting same branching behavior, maybe assert same output, or same termination behavior?

\section{Related work}\label{sec:related}

Murray et al \cite{murray2013} prove information flow security properties for the seL4 kernel, which is implemented in C.
Their proofs are on a more abstract level, and rely on previously established refinement proofs between the abstract level and the actual C code.
In contrast, our project allows fine-grained reasoning about the low-level layout of data using separation logic, which is more convenient for certain applications where the security property can only be stated by speaking about the low-level layout of memory buffers, i.e. in cases where abstracting into a higher-level language makes no sense.

Moreover, the information flow statement for seL4 is stated with respect to big-step operational semantics, which are not directly usable to reason about concurrency.
Equivalence to small-step semantics was therefore proved \cite{schirmer2006}, but recent success with concurrent separation logic \cite{appel2017} suggests that it is worth looking at a continuation-passing style small-step definition.

Beringer \cite{beringer2011} describes relational decomposition, a technique to reduce proofs involving two executions, such as non-interference proofs, to proofs involving only one execution.
This might turn out to be useful for the soundness proofs of our IFC Hoare rules, but it requires that one already has stated an information flow security property, so it solves a different problem than the main contribution of this paper.




Costanzo and Shao \cite{costanzo2014} prove non-interference for a simple imperative toy language
with pointer arithmetic and aliasing, also based on Separation Hoare Logic.
They define instrumented operational semantics, where each value is a tuple of a value and a sensitivity label, and prove theorems to go from the instrumented semantics to erased semantics and vice versa.
This seems to be a nice way to prove non-interference, but it is not clear how well instrumenting the operational semantics would scale to a real-world language like C.

Beaumont et al \citep{beaumont2016} present a cross-domain desktop compositor, a device allowing to use the same screen, keyboard and mouse for different machines containing data of different sensitivity classifications.
Formally verifying the correctness of their input/output dispatch code would be a typical use case for the system proposed in this paper.

Murray et al \cite{murray2016} propose a toy language system which also has value-dependent classifications, but they allow branching on high data and show how to refine programs such that the number of execution steps is preserved, which is needed for compositionality in a concurrent setting.

\section{Conclusions}\label{sec:conclusion}

In conclusion, it is rarely a good idea to include the same section three times in a paper, or to have a conclusion that does not conclude.

\appendix

%

\begin{acks}
TODO some source of money paid for Sam's visit in Melbourne.
\end{acks}

\bibliographystyle{ACM-Reference-Format}
\bibliography{IFC}

\end{document}